# A low density wave's spiral pattern speed, from the tracer separations (age gradient) across a spiral arm in the Milky Way

Jacques P Vallée

Herzberg Astronomy and Astrophysics, National Research Council of Canada

ADDRESS   5071 West Saanich Road, Victoria, British Columbia, Canada V9E 2E7

ORCID   *http://orcid.org/0000-0002-4833-4160*   EMAIL   jacques.p.vallee@gmail.com

**Abstract.** We observe the density wave's angular pattern speed $\Omega_p$ to be near 12-17 km/s/kpc, by the separation between a typical optical HII region (from the spiral arm's dust lane) and using a HII-evolution time model $T_{HII}$ to yield its relative speed, and independently by the separation between a typical radio maser (from the spiral arm's dust lane) with a maser model. The assumption of a fixed circular rotational speed of the gas and stars with galactic radius is employed (neglecting spiral perturbation at mid radii, nuclear bar influence at small radii, and tidal effects at large radii).

### 1. Introduction.

Spiral arms have been predicted by many theories, covering their origin, formation, and maintenance. For examples, a theory proposed that spiral arms are generated by the rotation of a bar located at the galactic center (predicting 2 or 4 spiral arms – Englmaier & Gerhard 1999), and another theory proposed that spiral arms are created by the passage of a nearby galaxy (predicting 2 tidal spiral arms). Some theories proposed that dynamic transient recurrent turbulences in the galactic disk would create spiral arms – see reviews in Dobbs & Baba (2014) and Pettitt et al (2020a). Dobbs & Baba (2014 – Section 3.7) proposed dynamic transient spiral arms with corotation everywhere (no gas flow through the arms, gas falling in an arm from both sides, gas can still undergo shocks as it falls into an arm – from both sides of each arm); unfortunately, a shock on the outside arm edge has seldom been observed (nor a dust lane there). These would resemble 'material arms'.

Another theory proposed that a triaxial dark matter halo could generate several spiral arms (Hu & Sijacki 2016). Yet another theory proposed that a spiral density wave, rotating at a fixed pattern speed, can create several spiral arms (see reviews in Shu 2016 and Roberts 1975). The density wave theory predicts that, over time, an arm tracer (a star, say) is born in a spiral arm, then it exits its spiral arm while going around its orbit around the Galactic Center, and thus we can infer the time $T_{arm}$ it takes to go from the inner arm edge (dust lane) to the outer arm edge (near 4 Myrs). Carrying on, that tracer reaches the next spiral arm, and thus we can infer the time $T_{next}$ from birth to reaching the next spiral arm (near 100 Myrs). That tracer now mixes with the distribution of the same tracer in that next spiral arm, making it difficult to see an age gradient for that tracer.

Each theory can be responsible for arms in some disk galaxies. Fortunately, observations of disk galaxies can check some of these theoretical predictions.

Only the density wave theory predicts (a) a *separation* in space for different arm tracers ('shock'/dust location versus the location of the 'potential minimum'/broad diffuse CO gas, say), predicts (b) an ordering of arm tracers along an 'age gradient' across the width of a spiral arm, and predicts (c) the age gradient to show a *reversal* across the Galactic Meridian (at galactic longitude of $0°$) – see reviews by Lin & Shu (1964); Roberts (1975); Gittins & Clarke (2004); Shu (2016).

These three specific predictions were observed in the Milky Way galaxy, beginning in Vallée (2014a). Letting a telescope drifts in galactic longitudes, along the Galactic plane in the disk of our galaxy, one can look across a specific spiral arm, in a specific tracer, and find the galactic longitude that is tangent to a spiral arm. Thus one measures the precise galactic longitude value when that arm tracer peaks in intensity – giving the location (galactic longitude) in that specific spiral arm. Each spiral arm consists of many diverse tracers (dust, masers, cold gas, etc). Selecting two different arm tracers should give an angular separation in galactic longitudes. With enough data for a given arm tracer, one sees each arm tracer's mean galactic longitude separating from the mean galactic longitude of another arm tracer, in the same spiral arm. A tracer age cannot be obtained for some tracers (dust, HI gas, relativistic electrons, say).

We have constructed an updated catalog, listing all arm tracers for each spiral arm segment; it doubled in length from the earlier catalog of Vallée 2016a; as a result, the typical relative error in tables 1 and 2 of that paper has now been improved substantially (in preparation).

Point (a) - separating arm tracers: one can assemble all these observational telescope scans, listing observed arm tangents (precise galactic longitudes) for each different arm tracer, for each different spiral arm, in a massive catalogue (Vallée 2016a), giving the mean galactic longitude where the tangent to each spiral arm is seen in each different tracer: cold $^{12}$CO gas (his Table 4), rare $^{13}$CO gas (his Table 5), ionized HII gas (his Table 6), maser spots (his Table 7), atomic HI gas (his Table 8), dust at 870 microns (his Table 9), old stars at 1 to 8 microns (his Table 10).

Observations showed that the gas and stars orbiting circularly around the galactic center have a fixed linear speed, to zero order (the speed may vary a little, due to perturbations on their trajectories from the Galactic bar and other spiral arms). This fixed value is observed over a large galactic radius range (typically from 3 to 30 kpc). Stars may also have some epicyclic motions (elliptical orbits), but we assume them to be negligible.

The linear spiral pattern speed of the density wave (fixed angular speed times galactic radius) differs from the fixed observed linear speed of the gas and stars in a circular orbit. When these two speeds are equal, then that galactic radius is said to be at 'corotation'. Inside that corotation radius, stars and gas go faster than the slower moving density wave pattern, creating a compression lane or even a 'shock' at the entrance to the arm, both compression or shock increasing the local star formation rate. This compression or 'shock' causes the bunching of dust into a 'dust lane', as well as the gravitational collapse of pockets of gas, forming protostars, with associated masers, etc. Protostars become ultracompact HII regions (masers) at radio wavelengths (0.2 pc size; 0.7 Myrs old – Xie et al 1996), and later become very young stars at optical wavelengths (1 pc; 1 Myrs), some with an optically-visible diffuse ionised HII regions (40 pc; 1.5 Myrs), then naked stars, and later young stellar open clusters (100 pc; 10 Myrs), and finally old stars and old star clusters. All are continuing in their circular orbit around the Galactic Center, leaving the shock and dust lane way behind, thus giving rise to the concept of an 'age gradient'.

If we measure the linear distance from the Sun to a spiral arm seen tangentially, then we can multiply this linear distance with the angular separation in galactic longitude between two arm tracers across that spiral arm. This gives us the linear separation of each arm tracer, across that spiral arm. For a given arm tracer, we can make a cross-cut of each spiral arm, then stack all these cross-cuts (one for each spiral arm) on top of each other, to yield a statistically merged cross-section encompassing all spiral arms. This gives us a much better precision for the relative linear separation for that arm tracer (in parsecs).

Point (b) - ordering arm tracers: we can compute a statistical mean cross-cut of a spiral arm, showing the mean position of each different arm tracer. We split all tracers into 4 colored zones, inside a cross-cut of a spiral arm. An age gradient shows up: the 'red' zone encompasses shock and a dust lane (at right); the 'orange' zone encompasses masers and young protostars; the 'green' zone encompasses young stars and thermal and relativistic electrons; and finally the 'blue' zone encompasses the potential minimum, old stars and the intensity of broad diffuse CO gas (in a beam width of 8'; not a gas overdensity). Here I wish to separate dense gas clouds near the masers and inner arm edge, from the broad diffuse gas near older stars and the outer arm edge.

Since then, the number of published observed telescope scans in galactic longitudes (measuring arm tangents in different tracers) went from 43 in 2014 to 125 in 2016, and to 204 in 2021, and yet tracers stayed in their proper colored zones; these 4 zones have been well confirmed with the addition of more telescope scans discovering the arm tangents, seen in many arm tracers: Vallée (2014a – Fig.1); Vallée (2014b – Fig.2); Vallée (2016a – Fig.2); Vallée (2017 – Fig.4); Vallée (2019a – Fig.3); Vallée (2020a – Fig.7); Vallée (2020b – Fig.11). Hou and Han (2015) obtained a roughly similar offset between gas and stars, but using much less telescope scans.

**Figure 1** shows the typical spiral arm width, going observationally at right from the inner arm edge (hot dust) to the center at the potential minimum (cold diffuse CO, old stars). Each arm tracer (vertical) is the mean separation of several telescope scans (for telescope scan details, see Vallée 2017 – table 2; Vallée 2014b – tables 3 to 10). The gas flow is from right (entering the dust lanes) to the left (exiting at the Potential minimum). This covers Galactic radii from 3 to 8 kpc (inside the sun"s galactic radius). $^{26}$Al is the arm tracer of the Aluminum 26 atom (see fig. 1 in Chen et al 1996). The density wave potential minimum is indicated by a vertical line (0 pc). The approximate extent of the spiral arm width is shown by a dashed horizontal line at bottom.

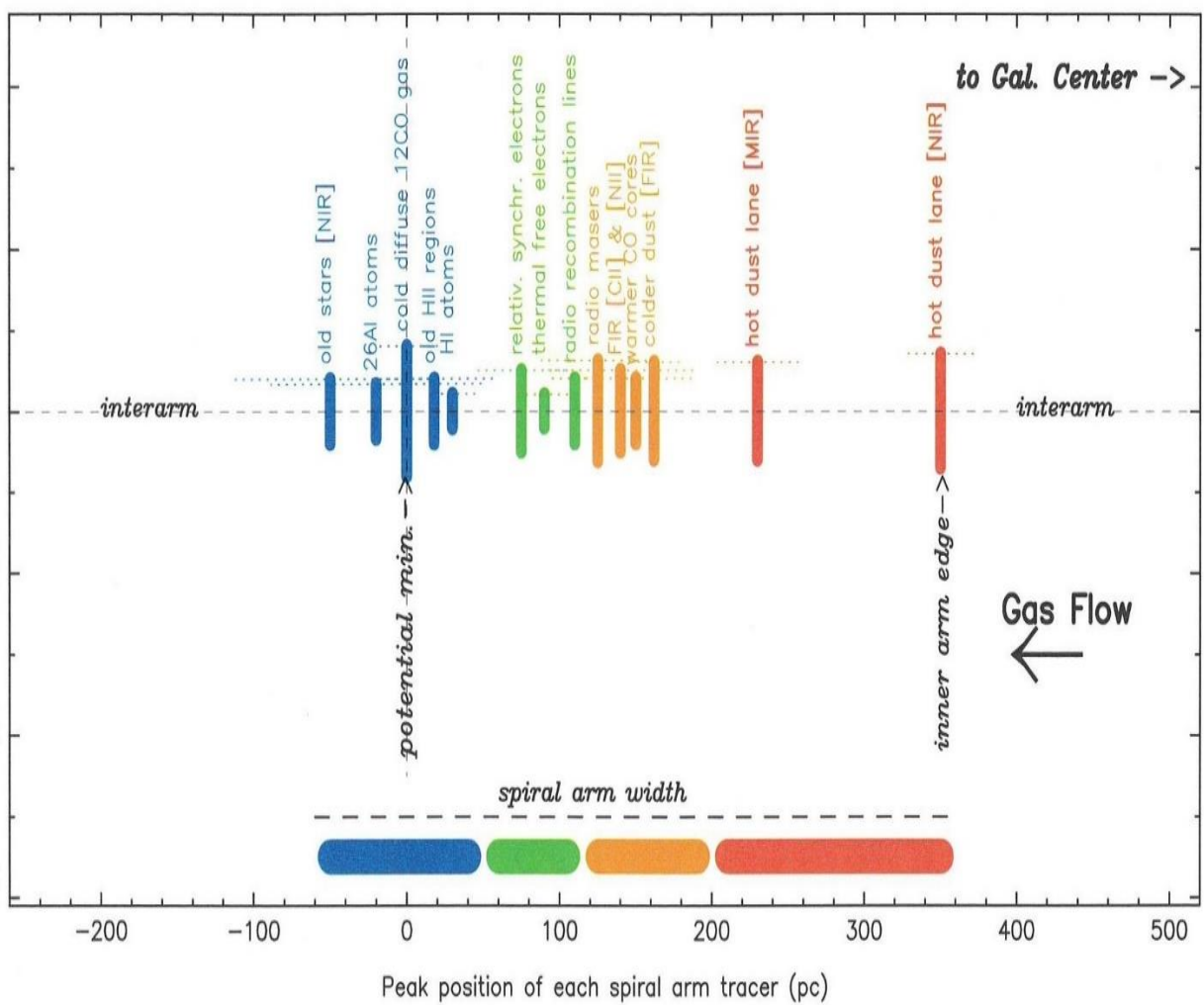

**Figure 1**. Schema of the mean observed position of each arm tracer, from its distance to the broad diffuse CO 1-0 gas position, as averaged over six spiral arm segments (Carina – near galactic longitude 283º, Crux-Centaurus near 310º, Norma near 328º, Perseus start near 338º, Scutum near 30º, Sagittarius near 50º). The y-axis is arbitrary, chosen to yield enough space for writing each tracer vertically. The origin of the x-coordinate, with the CO arm tracer, is set at 0 parsec. Four colored zones encompassed all diverse arm tracers: a 'red' zone (shock and dust lane); an 'orange' zone (masers and protostars); a 'green' zone (thermal and relativistic electrons, radio recombination lines from young stars); and a 'blue' zone (potential minimum, old stars and peak of broad diffuse old CO gas). Here 'old stars' is a very general term; the maximum surface density of old stars is normally associated with the local spiral potential minimum. This colored gradient is akin to an age gradient. The gas flow direction is shown (arrow, at right), as well as the direction to the Galactic Centre (top, right). The approximate width of the arm is shown (dashed horizontal line).

The shock is upstream from the spiral potential minimum (Gittins & Clark 2004). The statistical mean linear separation between the dust lane near the shock (density wave's 'potential maximum') and the broad diffuse CO 1-0 gas and concentration of old stars (density wave's 'potential minimum') is about **315** parsecs (Table 1 in Vallée 2016a, for an average dust location at several wavelengths: 2.4μm, 60 μm, 240 μm, 870μm separated from the CO peak), and about **365** parsecs (Table 2 in Vallée 2017, for an average 2.4μm dust peak separated from the CO peak; see Figure 1 here), when averaging over the Milky Way's inner spiral arms. These values are for Galactic radii between roughly 4.5 and 7.5 kpc, thus below corotation of gas and spiral pattern. Density wave theory predicts a slow decrease of these values with an increase in galactic radius, up to the co-rotation radius – see Section 2.

Our four colored zones (Fig. 1) are reminiscent of the prediction of three 'lanes' by Lin et al (1969, their Section 6), namely a dust 'lane' (our red zone), a 'lane' of brilliant young stars with HII regions (our green zone), and a 'lane' of dying stars (our blue zone). We have additionally inserted a maser and protostar 'lane' (our orange zone). We also provided the observed distances of each colored zones from each other, for the Milky Way galaxy: blue set at 0 pc, green observed at 80 pc, orange at 160 pc, red near 300 pc (extending to 365 pc – see Figure 1).

Separation and ordering will ensue for any kind of arm tracer for a pattern speed not equal to that of the local circular velocity of gas and stars. This would eliminate dynamic transient spiral arm models for our Galaxy.

Point (c) - reversal of arm tracers: observationally, one find a mirror-image symmetry going across the Galactic Meridian (longitude 0 degree): the red zone (dust) in each spiral arm is always closest to the Galactic Meridian, and the blue zone (old stars, concentration of broad diffuse CO gas) in each spiral arm is always farthest from the Galactic Meridian. The Galactic Meridian (galactic longitude 0°) separates the two facing sets - see Fig. 4(a) in Vallée (2017).

Stars and gas go in a roughly circular orbit around the Galactic Center at a fixed speed. In our Milky Way Galaxy, this fixed speed is $v_{gas}$ = 233.4 ±1.5 km/sec (Drimmel & Poggio 2018) over a very large range in Galactic radii (to zero order). This is a very crude approximation, there could be a systematic decrease of the rotational speed for small galactic radii below 5 kpc, and the spiral arms could inflect a localised perturbation in azimuth. Recent observational data put the Sun at a galactic radius of 8.1 kpc (Arbuter et al 2019). For comparison, these recent values slightly improve on the earlier review of Bland-Hawthorn & Gerhard (2016 – their fig.16), who had determined the Galactic rotation curve to undulate somewhat (less than 10%) between a radius of 5 kpc and 13 kpc (using $R_{sun}$ = 8.2 kpc a fixed $V_{gas}$ = 238 km/s).

At the Sun's distance to the Galactic Center, a value of 15 km/s/kpc gives a linear velocity of *122 km/sec*. This is smaller than the orbital velocity of the gas and stars (*233 km/s*), ensuring that a shock is created as the interarm gas enters a spiral arm.

For the Milky Way, the observed arm width (shock to potential minimum) and these different speeds yield a time elapsed $T_{arm}$ of $T_{arm}$ = **365 pc** / (233 – 122) km/s = **3.3 Myrs**. It follows that, to see an age gradient inside a spiral arm width, one must use arm tracers younger than 4 million years or so to cross the spiral arm width. For $\Omega_p$ = 12 km/s/kpc (see Eilers et al 2020), the linear pattern speed at the Sun's galactic radius is 97 km/s, and $T_{arm}$ becomes 2.7 Myrs.

With the Sun at a galactic radius of 8.1 kpc (Arbuter et al 2019), the time $T_{orbit}$ to do an orbital circle around the Galactic Center is $T_{orbit}$ = 2 π 8100 pc / 233 km/sec = **218 Millions of years**. But for a star at 5.0 kpc from the Galactic Center, the time to do a circle around the Galactic Center is smaller, or about $T_{orbit}$ = 135 Myrs.

This reversal of arm tracers around the Galactic Meridian (longitude 0°) can be explained as follows: for the longitudes l > 0°, one is looking at an arm from the downstream side of the gas flow (Figure 2), while for the longitudes l < 0⁰ one is looking at an arm from the upstream side, thus one sees a natural mirror-flip in the ordering of the arm tracers.

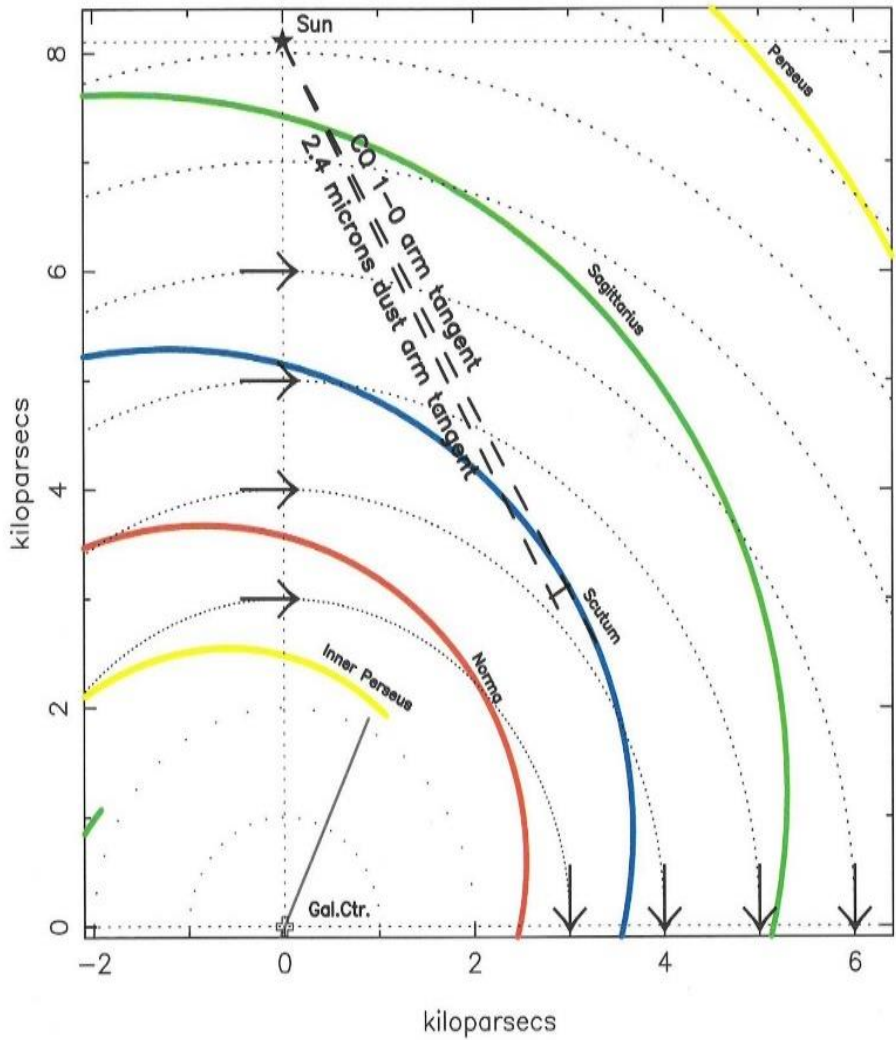

**Figure 2**. A view of Galactic quadrant I, with the gas flow direction (arrows). For the Scutum spiral arm, we plotted the arm tangents (dashed lines) for the CO 1-0 broad (hpbw 8') peak intensity, and for the 2.4 μm dust lane.

Elsewhere, several arm tracers have been observed in 24 nearby spiral galaxies. The angular separation among arm tracers and the pattern speed were also measured, albeit with less precision.

There, spiral arms showed a linear separation between an arm tracer in the starforming regions (dust, B-band, HI, etc) and an arm tracer in the old star regions (CO, optical star cluster, etc). The statistical median and mean linear separation between these two tracers was found statistically about **326** pc and 370 ±50 pc (Table 3 in Vallée 2020b). Also, we found that, for these 24 nearby disk galaxies with spiral arms, the median $T_{arm}$ was about **4.0 Myrs** and the $\Omega_p$ covered the range from 10 to 72 km/s/kpc – see Table 4 in Vallée (2020b).

For the two-armed disk galaxy UGC3825, Peterken et al (2019) compared the location of Hα near the dust/shock lane with the peak location of young stars (aged about 20 Myrs), and found a linear separation decreasing with the galactic radius (their Fig. 2); they repeated this comparison with old stars (about 200 Myrs) and found a similar pattern, measuring the *same* angular spiral pattern speed (their Suppl. Fig. 2).

For the spiral galaxy M51, Chandar et al (2017) compared the location of the dust lane (inner arm edge) and that of bright young stellar clusters (aged less than 6 Myrs) and found a separation of about **220 pc** (their Section 3.3); they found another separation of about 1 kpc (their Fig.8) between these star clusters (aged less than 6 Myrs) and the older star clusters (aged between 100 and 400 Myrs).

What is the value of the angular speed of the spiral pattern $\Omega_p$? Using radio maser data, the angular 'spiral pattern' speed $\Omega_p$ of the density wave theory appears to rotate ≤ 15 km/s/kpc (see Section 4 and Table 3 in Vallée 2020c for the Perseus and Cygnus arms), as based on the assumption that a spiral pattern can go to large distances from the Galactic Center.

*Some recent values of $\Omega_p$ made using optical stellar Gaia data were pretty high (near 30 km/s/kpc), and come from a different methodology (low range in galactic radius; using few stars in Sagittarius and Perseus arm). Fitting a galactic spiral density wave pattern $\Omega_p$ should include observational data gathered over a wider range of the galactic radius (as opposed to using only observational data near the Sun), and it should employ a 'log-spiral' mathematical shape as observed on the large scale (as opposed to using only neighbouring data that could also be fitted with a 'linear' arm shape); thus, fitting local Gaia stars and hoping that it should work over a much larger area is not a good solution (different local arm segments can yield very different global results, and different pitch angles). These arguments can explain that recent analyses employing local observed data (e.g. small galactic radial range, and stars from the local armlet) have come up with a high value for the pattern $\Omega_p$ near 30 km/s/kpc - see Dias et al (2019, who used polynomial arms and found $\Omega_p$ =28.2 km/s/kpc), Lépine et al (2017, who used objects in the local armlet and found $\Omega_p$ = 28.5 km/s/kpc), Michtchenko et al (2018, using nearby stellar Gaia data and finding $\Omega_p$ = 28.0 km/s/kpc), and Grosbol & Carraro (2018, using B and A stars, finding $\Omega_p$ between 20 and 30 km/s/kpc). The nearby Local Arm is too short and twisted to be included as a density wave's global log-spiral arm so its stars should not be included in determining $\Omega_p$; if included, interarm objects would bend the true arm shape and its true pitch angle. The Local Arm is an 'interarm island' near the Sun – it does not show the interarm tracers to be ordered, as in Figure 1 here (see Vallée 2018). Other large interarm features or interarm islands have been found between the Perseus and Cygnus arms (Vallée 2020c) and beyond the Cygnus arm (Table A2 in Koo et al 2017). Some short interarm spurs can be found near long arms (Vallée 2020c) and sometimes they may be misconstrued as a deviation of a real arm; hence it is appropriate to look in both Galactic Quadrants I and IV to find a single long real arm and then fit a single log-spiral to both segments (Fig. 1 in Vallée 2015).*

Observed masers and CO locations indicate that the Perseus arm and the Cygnus arm have masers observed in front of their stars, thus ensuring a higher value for the galactic corotation radius, beyond 15 kpc (Fig. 1 in Vallée 2019b), and ensuring a lower value for $\Omega_p$.

Studies of the bar(s) near the Galactic Center indicated several bars – the 0.4kpc-radius 'nuclear bar' (Table 2 in Vallée 2016b), the 2.1-kpc radius short 'boxy' bar (Table 3 in Vallée 2016b), and the 4.2-kpc radius 'long' bar (Table 3 in Vallée 2016b), all based on observations. Many interpretations of these observations led to some preferring the 'long' thin bar (Section 4.3 and Fig.9 in Bland-Hawthorn & Gerhard 2016) or the boxy 'short' bar (Fig. 3 and section 3.2 in Vallée 2016b). It could be that the 'long' thin bar may be a composite of segments, apparently crossing some long spiral arms but not affecting them (see the inner Sagittarius arm at l=343º and at a radius of 2.4 kpc; see the inner Norma arm at l =+20º and at a radius of 2.8 kpc). The 'long' thin bar, if physically real, would stop the spiral arms. Hence here, we adopt the boxy 'short' bar, which does not physically cross the inner spiral arms, and stays well away from the inner arm tangents.

In what follows, we seek the various timescales, and the angular pattern speed of the density wave $\Omega_p$ using data at different galactic radius.

## 2. New results

Here we will employed the separation between some arm tracers, and an appropriate time model, to calculate the pattern speed in the Galactic disk.

**Table 1** shows recent values of the angular pattern speed $\Omega_p$, from the recent observational literature, and their resulting consequences.

At what distance is the 'corotation' radius ? For an angular pattern speed of 15 km/s/kpc, the corotation radius is 233/15 = 16 kpc; this would fall past the Cygnus arm, in the Milky Way disk – see Fig. 1 in Vallée (2020c). At $\Omega_p$ = 11, one gets a corotation radius near 21 kpc.

**Table 2** shows our computed results for the linear pattern speed $v_p$, the relative velocity ($v_{gas} – v_p$), a linear tracer separation, for each of three values for the density wave angular pattern speed (10, 12, and 15 km/s/kpc), at a time of 2 (col.4) and 4 (col.5) Myrs. At the bottom of the table at right, one sees that the pattern speed overcomes the speed of the gas orbiting around the Galactic Center, so the relative velocity changes sign. When the two speeds are equal, the relative speed is zero, then we are at the corotation radius.

2.1 Optical HII regions as an arm tracer

For a specific arm tracer, in the age gradient across a spiral arm, the observation of its distance S from the inner arm's dust lane and its age T from a model (based on its linear extent) yield its relative linear speed S/T, and then subtracting that from the fixed model orbital gas speed around the Galactic Center (see Introduction above) gives the density wave's linear spiral pattern speed, which will become angular by dividing by the galactic radius of that arm's segment.

Optically detected Galactic HII regions have been compiled, with their angular and linear diameters, and their distribution with diameter has been obtained. Ismael et al (2005 – their Fig. 1d) found an exponential decrease of the number of HII regions while increasing the linear diameter; a diameter value near 40 pc seems typical of a mid-range HII region.

Averaging over 6 half-arms (Carina, Crux, Norma, Perseus start, Scutum, Sagittarius), Vallée (2017 – his Table 2) showed an average observed 59 pc separation of HII complexes from the broad mean CO longitudes, and an average observed 318 pc separation of the 'hot' (2.4 to 60 µm) dust from the broad CO longitude, giving by subtraction an average observed 259 pc separation of HII complexes from the hot dust lane.

As a first example, optically-visible HII regions (typical diameter of 40 pc) have an age $T_{HII}$ near **1.5 Myrs** (Fig. 6e and Fig.7 in Hunt & Hirashita 2009), assuming extragalactic HII regions follow the same size versus density relation as the one in our Galaxy but with different initial conditions, limited by dust absorption. These HII regions seen along a tangent to a spiral arm at a mean galactic radius $R_{Gal}$ of 5 kpc are separated from the combined (2.4µm and 60µm) hot dust lane by $S_{HII}$ = 259 pc (see above); then we can write:

$$S_{HII} / T_{HII} = v_{gas} - (R_{gal} \cdot \Omega_p) \qquad \ldots \text{Equ. (1)}$$

with the known circular orbital gas speed $v_{gas}$ of 233 km/s, this equation yields a value for $\Omega_p$ = 12.1 km/s/kpc. This is an average over 6 arm segments over Galactic Quadrants IV and I. Included arm segments are already listed in the caption to Figure 1; excluded arm segments are those with an arm tangent too close to the Galactic Center, having too few HII data available yet (Start of Sagittarius near galactic longitude 346°, Start of Norma near 018°). Owing to uncertainties, the individual points in Figure 3 can be up or down from that mean value. One is not trying to fit each individual point for each spiral arm. A rough estimate of the uncertainty in this value can be had by recalculating the same typical HII regions, with this time a $T_{HII}$ of 1.6 Myrs, yielding a value for $\Omega_p$ near 14.2 km/s/kpc. Uncertainties in the gas speed (±1.5 km/s) and in the offset (±50 pc) must be included, giving a combined error of ±3 km/s/kpc.

**Figure 3** shows in vertical coordinates the separation, in each spiral arm, between the shock (dust lane, maser lane) and the Potential Minimum (broad diffuse CO lane peaking in intensity at a specific galactic longitude), expressed in parsecs, as a function of the galactic radius of the measurement. The black slanted line shows the predicted separation between the shock and the Potential Minimum after an elapsed time of 1.5 million years, for an angular pattern speed $\Omega_p$ of 10 (dashes) or 15 (dots) km/sec/kpc. Dust absorption along the line of sight from Earth makes it difficult to observe optical HII regions at small galactic radii; the expansion of an HII region with time makes it difficult to see much older optical HII regions.

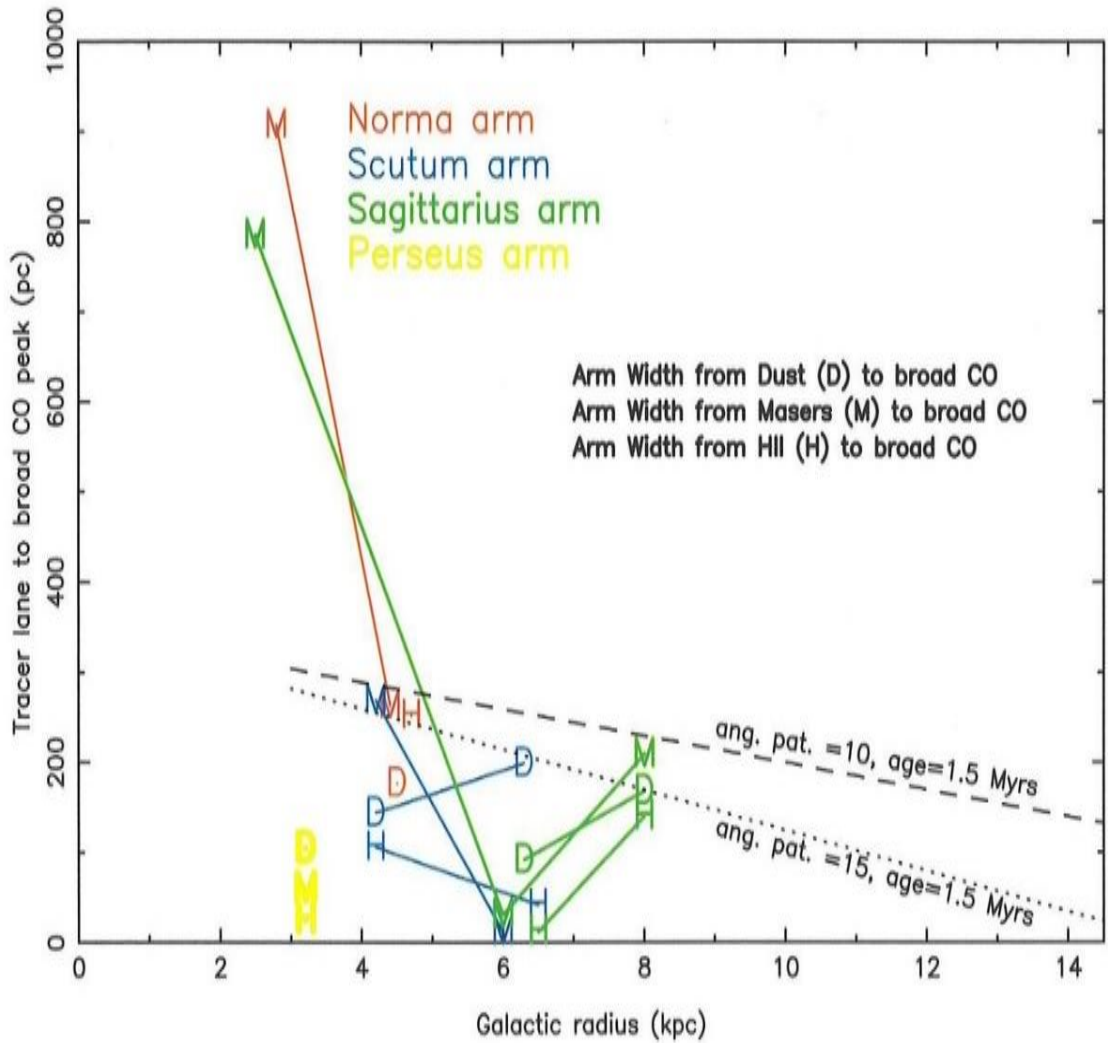

**Figure 3.** For each spiral arm, the vertical axis gives the separation between a tracer longitude (D=dust; M=maser; H=HII complex) and the broad diffuse CO tracer longitude, as a function of the galactic radius (horizontal axis). Data from observed tangents to spiral arms – see Table 3. The predictions for an angular speed pattern $\Omega_p$ are shown for 10 (dashes) and 15 (dots) km/s/kpc, after 1.5 Myrs (pertinent for typical HII complexes observed at optical wavelengths from Earth). The dotted and dashed lines meet the measurements of the HII complex (H) data near 260 pc at moderate galactic radii.

In this figure, the two maximum HII offsets (in Norma at 262 pc and 4.5 kpc; in Sag.-Carina at 140 pc and 8.0 kpc) are close to the range of $\Omega_p$ between 10 (dashes) and 15 (dots) km/s/kpc, while in other arms segments the offsets appear smaller, possibly due to dust absorption inhomogeneities at optical wavelengths in different line-of-sights and at different distances to the Sun. Dust may be causing the offsets of the HII regions to change at larger Galactic radii as dust density and metallicity decreases.

The $\Omega_p$ result from Equation (1) above should be more secure.

2.2 Shock location (dust / maser lane) versus the location of the 'Potential Minimum' of the density wave (broad diffuse CO J=1-0 / old stars)

**Table 3** shows the observed locations of dust lanes, masers, and HII complexes in the Milky Way's spiral arms, as obtained from telescope scans in galactic longitudes.

Radio detected Galactic compact masers have been compiled, Averaging over 4 half-arms (Norma, Perseus start, Scutum, Sagittarius), Vallée (2017 – his Table 2) showed an average observed 201 pc separation of masers from the broad mean CO longitudes, and an average observed 318 pc separation of the 'hot' (2.4 to 60 µm) dust from the broad CO longitude, giving by subtraction an average observed 117 pc separation of masers from the hot dust lane.

As a second example, with an age near 0.7 Myrs, these maser regions seen along a tangent to a spiral arm at a mean galactic radius $R_{Gal}$ of 5 kpc are separated from the combined (2.4µm and 60µm) hot dust lane by $S_{HII}$ = 117 pc (see above); then we can use Equ. 1: With the known circular orbital gas speed $v_{gas}$ of 233 km/s, this equation yields a value for $\Omega_p$ = 13.2 km/s/kpc. This is an average over 4 arm segments over Galactic Quadrants IV and I. Included arm segments are listed just above (Norma, Perseus start, Scutum, Sagittarius); excluded arm segments are Carina and Crux-Centaurus (too few maser data available yet). Again, here one is not trying to fit each individual point for each spiral arm. A rough estimate of the uncertainty in this value can be had by recalculating the same typical maser regions, with this time a T of 0.8 Myrs, yielding a value for $\Omega_p$ near 17.4 km/s/kpc. Uncertainties in the gas speed (±1.5 km/s) and in the offset (±50 pc) must be included, giving a combined error of ±3 km/s/kpc.

What is the time for a tracer to reach the next spiral arm along the tracer's circular orbit $T_{next}$? Over a longer time, tracers would have time to wander away from their initial arm, and reach the next spiral arm (confusing the locations there of the same tracers born in that next arm). For $\Omega_p$ = 15 km/s/kpc, and to reach the next spiral arm, away at 0.25x 2 π 8.1 kpc = 12.7 kpc, at a relative speed of (233 - 122) = 111 km/s, a star needs a time $T_{next}$ = 114 Myrs. It could be less or more, depending on the tracer's own higher or smaller peculiar velocity. Thus 4 times that amount gives the time to return inside the original spiral arm $T_{return}$ = 456 Myrs.

If the value of the angular pattern speed is $\Omega_p$ =10, then the time for a star to reach the next spiral arm is reduced to $T_{next}$ = 84 Myrs (see Table 1), so then 4 times more yields $T_{return}$ = 336 Myrs.

It could then be confusing to an observatory, looking in that next arm, to see both a tracer made in its own arm and that same tracer made in the preceding arm. It follows that the maximum time to see an age gradient <u>inside a spiral arm</u> width is roughly $T_{arm}$ = 4 Myrs (see Fig.1 and Section 2), and the maximum time to follow an age gradient <u>in the first interarm</u> is around $T_{next}$ = 99 Myrs (mid value between $\Omega_p$ = 15 and 10). In addition, stars formed in the inter arm regions (e.g. in the local arm) may also be a source of confusion.

What is the maximum time to see an age gradient $T_{max}$ in a tracer? Some arm tracer, like open or globular stellar clusters, can have enough gas to allow new star formation inside their border. These new stars could confuse the measurement of the elapsed time since the formation of the oldest stars in that stellar cluster. If the stellar cluster was formed in the shocked dusty lane at the inner edge of a spiral arm, then only the oldest stars in that cluster should be used for determining the cluster's age. Bursts of star formations nearby, occurring every **20** Myrs or so (Comeron et al 2021), could lower the average age of all stars in that cluster. Also, several star forming regions have been reported to have a continuous formation of star over a period of 5-10 Myr.

Preliminary comparisons with Gaia observations can be made here. The Gaia satellite has offered us many catalogs of nearby stars at optical wavelengths, and their locations near spiral arms in the Galactic plane. These images show a diffuse broad area for the Perseus arm and for the Sagittarius arm. On these images we can see the well-defined narrow arm 'spine' indicated by the radio masers: see Fig.1 in Kounkel et al (2020); Fig.3 in Khoperskov et al (2020); Fig. 8 in Cantat-Gaudin et al (2020), Fig. 12 in Ferreira et al (2020). From the open star cluster distributions, the arm width appears to be large. For example, using the Gaia catalogue, the optical star clusters with an age between **10 and 50** Myrs, identified in Kounkel et al (2020 – their Fig.1) are distributed along the Perseus and Sagittarius spiral arms, but the star clusters with an age of 200 to 630 Myrs do not show these spiral arms. Also using the Gaia catalogue, open clusters with an age below 158 Myrs do not appear to show the Perseus and Sagittarius spiral arm (Figure 8.2 in Cantat-Gaudin et al, 2020); hence a cut at a much younger age is needed to show the spiral structure. To summarize, the open clusters needed to see an age gradient must not be older than $T_{max}$ = **50** Myrs or so.

### 3. Conclusion

We found a rough angular spiral pattern speed $\Omega_p$ near 12-17 km/s/kpc in two ways: first, using a mean separation of a typical optical HII region from the dust lane of 259 pc, at a galactic radius near 5 kpc (Section 2.1 and Equation (1)), and second, using a mean separation of a typical radio maser from the dust lane, at a galactic radius near 5 kpc (Section 2.2 with Equ.(1)). Both estimates employed a circular orbital speed of 233 km/s for the gas going around the Galactic Center; this rough value may be amended with newer data in the future for lower galactic radii.

*This low $\Omega_p$ value is similar to the value of 12 km/sec/kpc as obtained recently using luminous red giant stars (Eilers et al 2020). Smoothed particle hydrodynamics by Pettitt et al (2020b) employed a low spiral pattern speed near 20 km/s/kpc – they find arm tangencies to be well traced by young stars. Hydrodynamical simulations by Li et al (2016) had predicted a low spiral pattern speed near 23 km/s/kpc, when trying to match data from red clump giants in the Galactic Bulge. Analytical models by Lin et al (1969 – their Section 8b) had predicted a low pattern speed near 11-13 km/s/kpc, when comparing with young stars. Recent observational data argues for a value of near 15 km/sec/kpc or less, based on the location of radio masers 'in front' of the molecular clouds in the Cygnus arm (Fig. 1 and Section 4 in Vallée 2020c). Also, using radio HI data, $\Omega_p$ was found by Foster & Cooper (2010- their Section 3) around 20 km/s/kpc in the inner Galaxy, and around 16 km/s/kpc in the outer Galaxy. Using an overdensity of optical stars, $\Omega_p$ was found by Monguio et al (2015) around 16-20 km/s/kpc around the Perseus arm, placing that arm inside the co-rotation radius.*

Some authors (Foster and Cooper 2010, etc) argued for a varying $\Omega_p$ value with increasing Galactic radius, based on differing local conditions (higher $\Omega_p$ near the nuclear bar; lower $\Omega_p$ far from the Galactic nucleus due to distant tidal effects, etc).

Such a low value implies that aging stars orbiting the Galactic Center may leave their natal spiral arm and reach the next spiral arm in $T_{next}$ of about 100 Myrs.

Our time model, for the density wave's angular pattern speed, employed here is given in Table 2.

Here we computed various time scales, pertinent to spiral arms. Statistical analyses of the observational data showed a maximum time between 3 and 4 Myrs in our Galaxy (Section 2 and Fig.1) to cross the width of a spiral arm, while for 24 other spiral galaxies the mean maximum time $T_{arm}$ was found to be 4.0 Myrs. That seems to be the time $T_{arm}$ needed to see an arm gradient inside an arm width, when using the arm's own tracers.

**Acknowledgements.** The figure production made use of the PGPLOT software at NRC Canada in Victoria. I thank the referee for a thorough reading of the manuscript, as well as numerous enlightening insights that improved it, and correcting an oversight.

**Data availability.** All data underlying this article are available in the article, and/or will be shared on reasonable request to the corresponding author.

**References**

Abuter, R., Amorim, A., et al 2019, A geometric distance measurement to the Galactic Center black hole with 0.3% uncertainty. Astron. & Astrophys., v. 625, art. L10, pp.1-10.

Bland-Hawthorn, J., Gerhard, O. 2016, The Galaxy in context: structural, kinematic, and integrated properties. Ann Rev As Ap, v54, p.529-596.

Cantat-Gaudin, T., Anders, F., Castro-Ginard, A., et al. 2020. Painting a portrait of the Galactic disk with its stellar clusters. Astron. & Astroph., v.640, art.1, p.1-20.

Chandar, R., Chien, L.-H, Meidt, S., et al. 2017. Clues to the formation of spiral structure in M51 from the ages and locations of star clusters. ApJ, v.845, art. 78, p.1-12.

Chen, W., Gehrels, N., Diehl, R., Hartmann, D. 1996, On the spiral arm interpretation of Comptel $^{26}$Al map features.A&A SS, v120, p315-316.

Comeron, F., Djupvik, A.A., Schneider, N., Pasquali, A. 2021. The historical record of massive star formation in Cygnus. MNRAS, v644, a62, p1-13.

Dias, W.S., Monteiro, H., et al. 2019. The spiral pattern rotation speed of the Galaxy and the corotation radius with GAIA DR2. MNRAS, v486, p5726-5736.

Dobbs, C., Baba, J. 2014. Dawes review 4: spiral structures in disk galaxies. Publ Astron Soc Australia, v..31, art. 35, p.1-40.

Drimmel, R., Poggio, E. 2018. On the Solar velocity. Research Notes AAS, v..2, art.210, pp.1-5.

Eilers, A.-C., Hogg, D.W., Rix, H.-W., et al. 2020. The strength of the dyamical spiral perturbation in the Gaseous disk. ApJ, v.900, art.186, p.1-11.

Englmaier, P., & Gerhard, O. 1999. Gas dynamics and large-scale morphology of the Milky Way galaxy. MNRAS, v304, p512-534.

Ferreira, F.A., Corradi, W.J., Maia, F.F., Angelo, M.S. Santos, J.F. 2020. Discovery and astrophysical properties of galactic open clusters in dense stellar fields using Gaia DR2. MNRAS, v.496, p.2021-2038.

Foster, T., Cooper, B. 2010. Structure and dynamics of the Milky Way: the evolving picture. ASP Confer

Ser., 438, p16-30. In: 'The Dynamic Interstellar Medium: a celebration of the Canadian Galactic plane survey'; edited by R. Kothes et al.; Astron. Soc. Pacific, San Francisco, USA.

Gittins, D.M., Clarke, C.J. 2004. Constraining corotation from shocks in tightly wound spiral galaxies. MNRAS, v.349, p.909-921.

Grosbol, P., Carraro, G. 2018. The spiral potential of the Milky Way. A&A, v619, a50, p1-10.

Hou, L.G., Han, J.L. 2015, Offset between stellar spiral arms and gas arms of the Milky Way. MRAS, v454, p626-636.

Hu, S., Sijacki, D. 2016, Stellar spiral structures in triaxial dark matter haloes. MNRAS, v461, p2789-2808.

Hunt, L.K., Hirashita, H. 2009. The size-density relation of extragalactic HII regions. Astron. & Astrophys., v. 507, p.1327-1343.

Ismael,H.A., Alawy, A.E.,Takey, A.S., et al. 2005. Frequency distribution of the HII regions radii as a distance indicator. J. Korean Astr. Soc., v.38, p.7-12

Khoperskov, S. ,Gerhard, O., Di Matteo, P., et al 2020. Hic Sunt Dracones: Cartography of the Milky Way spiral arms and bar resonances with Gaia Data Release 2. Astron. & Astrophys., v.634, L8, p.1-10.

Koo, B.-C., Park, G., Kim, W.-T., Lee, M.-G., et al 2017, Tracing the spiral structure of the Outer Milky Way with dense atomic hydrogen gas. PASP, 129, 094102 (1-19).

Kounkel,M., Covey, K., Stassun,K. 2020. Untangling the Galaxy: II. Structure within 3 kpc. Astron. J., v.160, a.279, p1-22.

Lépine, J.R., Michtchenko, T., Barros, D., Vieira, R. 2017. The dynamical origin of the Local arm and the Sun's trapped orbit. ApJ, v843, a48, p1-12.

Li, Z., Gerhard, O., Shen, J., et al 2016, Gas dynamics in the Milky Way: a low pattern speed model. Ap J, 824, 13, p1-11.

Lin, C.C., Shu, F.H. 1964. On the spiral structure of disk galaxies. Ap J, v.140, pp.646-655.

Lin, C.C., Yuan, C., Shu, F.H. 1969. On the spiral structure of disk galaxies. III. Comparison with observations. ApJ, v.155, p721-746.

Michtchenko, T.A., Lépine, R.D., et al 2018. On the stellar velocity distribution in the solar neighbouhood in light of Gaia DR2. ApJ Lett.,v 863, L37, p1-6.

Monguio,M., Grosbol, P., Figueras, F. 2015. First detection of the field star overdensity in the Perseus arm. A&A, 577, A142, p1-11.

Peterken, T.G., Merrifield, M.R., Aragon-Salamanca, A., et al. 2019. A direct test of density wave theory in a grand-design spiral galaxy. Nature Astr., v.3, p.178-182.

Pettitt, A., Dobbs, C., Baba, J., et al. 2020a. How do different spiral arm models impact the ISM and GMC population ? MNRAS, v. 498, p.1159-1174.

Pettitt, A., Ragan, S., Smith, M. 2020b, Young stars as tracers of a barred-spiral Milky Way. MNRAS, 491. 2162-2179.

Roberts, W.W. 1975, Theoretical aspects of Galactic research.Vistas in Astronomy, v.19, p.91-109.

Shu, F.H. 2016. Six decades of spiral density wave theory. Annual Rev Astron & Astrophys, v. 54, p.667-724.

Vallée, J.P. 2014a. The spiral arms of the Milky Way: the relative location of each different arm tracer, within a typical spiral arm width. Astron J., v.148, art.5, p.1-9.

Vallée, J.P. 2014b. Catalog of observed tangents to the spiral arms in the Milky Way galaxy. ApJ Suppl.

Ser., v.215, art.1, p1-9.

Vallée, J.P. 2015. Different studies of the global pitch angle of the Milky Way's spiral arms. MNRAS, v450, p4277-4284.

Vallée, J.P. 2016a. A substructure inside spiral arms, and a mirror image across the galactic meridian. Astrophys J., v.821, art.53, p.1-12.

Vallée, J.P. 2016b, The start of the Sagittarius spiral arm (Sagittarius origin) and the start of the Norma arm (Norma origin): model-computed and observed at tangents at galactic longitudes $-20° < l < +23°$. Astron. J., v151, a.55, p.1-16.

Vallée, J.P. 2017. A guided map to the spiral arms in the galactic disk of the Milky Way. Astronomical Review, v.13, p.113-146.

Vallée, J.P. 2018. Meta-analysis from different tracers of the small Local Arm around the Sun – extent, shape, pitch, origin. Ap Sp Sci, 363, 243, p.1-9.

Vallée, J.P. 2019a. Spatial offsets between interstellar bone-like filaments, radio masers, and cold diffuse CO gas in the Scutum spiral arm. Ap Sp Sci, v.364, art.150, p.1-9.

Vallée, J.P. 2019b. Spatial and velocity offsets of Galactic masers from the centres of spiral arms. MNRAS, 489, 2819-2829.

Vallée, J.P. 2020a, A new multitracer approach to defining the spiral arm width in the Milky Way. Astrophys. Journal, v. 896, art. 19, p. 1-10.

Vallée, J.P. 2020b, Statistics on 24 spiral galaxies, having different observed arm locations, using different arm tracers. New Astron., v.76, art.101337, p.1-13.

Vallée, J.P. 2020c, Interarm islands in the Milky Way – the one near the Cygnus spiral arm. MNRAS, v. 494, p.1134-1142.

Xie, T., Mundy, L., Vogel, S., Hofner, P. 1996, On turbulent pressure confinement of ultra-compact HII regions. ApJ, 473, L131-L134.

**Table 1. Possible angular pattern speed of the density wave, and its effect on other physical parameters.**

| (1) Angular pattern speed of wave (km/sec/kpc) | (2) Linear pattern speed at Sun (km/sec) | (3) Relative speed (km/sec) | (4) Time for a star to reach the next arm | (5) Co-rotation radius (Myears) | (6) Comments on the Location of the corotation radius (kpc) |
|---|---|---|---|---|---|
| 10 | 81 | 152 | 84 | 23 | past the Outer-Scutum arm |
| 12 | 97 | 136 | 93 | 19 | before the Outer-Scutum arm |
| 15 | 122 | 111 | 114 | 16 | past the Cygnus (Outer-Norma) arm |
| 20 | 162 | 71 | 179 | 12 | past the Perseus arm |
| 25 | 202 | 31 | 410 | 9 | past the Sun's galactic orbit |
| 28.8 | 233 | 0 | infinity | 8.1 | at the orbit of the Sun around the Galactic Center |

Notes.
Column 1 gives the angular spiral pattern speed $\Omega_p$ as proposed in the recent literature, for the Milky Way.
Column 2 gives the linear pattern speed at the Sun's distance $R_{sun}$, through the equation: $R_{SUN} \times \Omega_p$
Column 3 gives the relative linear velocity between the gas speed at the Sun's location (233 km/s) and the wave's linear pattern speed.
Column 4 gives the time for a star to leave its spiral arm and to join the next spiral arm: $0.25 \times 2\pi R_{sun}$ / relative linear speed.
Column 5 gives the corotation radius, using the observed gas and star speed: 233 km/s / $\Omega_p$
Column 6 notes the location of the corotation radius, along the Galactic Meridian (Galactic longitude zero), relative to the spiral arms.

- - - - - - - - - - -

**Table 2. Age Gradient: linear separation of a tracer of age T (from the dust lane), for different angular pattern speed.**

| (1) Galactic Radius (kpc) | (2) Linear pattern speed $v_p$ (km/s) | | | (3) Relative ($v_{gas} - v_p$) speed 233 - $v_p$ (km/s) | | | (4) Linear Separation, from hot dust lane, at T=2 Myrs (parsecs) | | | (5) Linear separation, from hot dust lane, at T=4 Myrs (parsecs) | | |
|---|---|---|---|---|---|---|---|---|---|---|---|---|
| | $\Omega_p$=10 | 12 | 15 | $\Omega_p$=10 | 12 | 15 | $\Omega_p$=10 | 12 | 15 | $\Omega_p$=10 | 12 | 15 |
| 4 | 40 | 48 | 60 | 193 | 185 | 173 | 386 | 370 | 346 | 772 | 740 | 692 |
| 5 | 50 | 60 | 75 | 183 | 173 | 158 | 366 | 346 | 316 | 732 | 692 | 632 |
| 6 | 60 | 72 | 90 | 173 | 161 | 143 | 346 | 322 | 286 | 692 | 644 | 572 |
| 7 | 70 | 84 | 105 | 163 | 149 | 128 | 326 | 298 | 256 | 652 | 596 | 512 |
| 8 | 80 | 96 | 120 | 153 | 137 | 113 | 306 | 274 | 226 | 612 | 548 | 452 |
| 9 | 90 | 108 | 135 | 143 | 125 | 98 | 286 | 250 | 196 | 572 | 500 | 392 |
| 10 | 100 | 120 | 150 | 133 | 113 | 83 | 266 | 226 | 166 | 532 | 452 | 332 |
| 11 | 110 | 132 | 165 | 123 | 101 | 68 | 246 | 202 | 136 | 492 | 404 | 272 |
| 12 | 120 | 144 | 180 | 113 | 89 | 53 | 226 | 178 | 106 | 452 | 356 | 212 |
| 13 | 130 | 156 | 195 | 103 | 77 | 38 | 206 | 154 | 76 | 412 | 308 | 152 |
| 14 | 140 | 168 | 210 | 93 | 65 | 23 | 186 | 130 | 46 | 372 | 260 | 92 |
| 15 | 150 | 189 | 225 | 83 | 44 | 8 | 166 | 88 | 16 | 332 | 176 | 32 |
| 16 | 160 | 192 | 240 | 73 | 41 | -7 | 146 | 82 | n.a. | 292 | 164 | n.a. |
| 17 | 170 | 204 | 255 | 63 | 29 | -22 | 126 | 58 | n.a. | 252 | 116 | n.a. |
| 18 | 180 | 216 | 270 | 53 | 17 | -37 | 106 | 34 | n.a. | 213 | 68 | n.a. |
| 19 | 190 | 228 | 285 | 43 | 5 | -52 | 86 | 10 | n.a. | 172 | 20 | n.a. |
| 20 | 200 | 240 | 300 | 33 | -7 | -67 | 66 | n.a. | n.a. | 132 | n.a. | n.a. |
| 21 | 210 | 252 | 315 | 23 | -19 | -82 | 46 | n.a. | n.a. | 92 | n.a. | n.a. |
| 22 | 220 | 264 | 330 | 13 | -31 | -97 | 26 | n.a. | n.a. | 52 | n.a. | n.a. |
| 23 | 230 | 276 | 345 | 3 | -43 | -112 | 6 | n.a. | n.a. | 12 | n.a. | n.a. |
| 24 | 240 | 288 | 360 | -7 | -55 | -127 | n.a. | n.a. | n.a. | n.a. | n.a. | n.a. |

Note 1: owing to a tracer's peculiar velocity, the linear separation could differ from the values in columns 4 and 5.
Note 2: The fixed circular velocity may vary a bit with radius. Still, larger galactic radii are kept, to give much needed approximate values. The approximation decreases and the accuracy increases nearer the Sun's orbit.

- - - - - - - - - - - - -

**Table 3 - Offset between the arm tangent longitude for a tracer X, and the arm tangent longitude for a tracer Y** [a] [b]

| Arm name | X tracer Tangent Longit. (o) | Y=broad CO gas Tangent Longit. (o) | Ang. offset (o) | Linear offset at solar distance (pc) | Located at Galactic radius (kpc) |
|---|---|---|---|---|---|
| X= dust lane at 870μm | | | | | |
| Carina | 284.2 | 281.8 | 2.4 | 168 at 4.0 kpc | 8.0 |
| Crux | 311.4 | 309.5 | 1.9 | 199 at 6.0 kpc | 6.3 |
| Norma | 329.6 | 328.4 | 1.2 | 177 at 7.5 kpc | 4.5 |
| Perseus Start | 337.8 | 337.0 | 0.8 | 105 at 7.5 kpc | 3.2 |
| Scutum | 030.9 | 032.4 | 1.5 | 144 at 5.5 kpc | 4.2 |
| Sagittarius | 049.1 | 050.7 | 1.6 | 092 at 3.3 kpc | 6.3 |
| | | | | | |
| X= methanol maser | | | | | |
| Carina | 284.8 | 281.8 | 3.0 | 209 at 4.0 kpc | 8.0 |
| Crux | 309.6 | 309.5 | 0.1 | 11 at 6.0 kpc | 6.3 |
| Norma | 330.4 | 328.4 | 2.0 | 262 at 7.5 kpc | 4.5 |
| Perseus Start | 337.3 | 337.0 | 0.3 | 039 at 7.5 kpc | 3.2 |
| Sagittarius Start | 348.0 | 342.0 | 6.0 | 785 at 7.5 kpc | 2.5 |
| Norma Start | 015.5 | 023.5 | 8.0 | 907 at 6.5 kpc | 2.8 |
| Scutum | 029.6 | 032.4 | 2.8 | 269 at 5.5 kpc | 4.2 |
| Sagittarius | 050.2 | 050.7 | 0.5 | 029 at 3.3 kpc | 6.3 |
| | | | | | |
| X= HII complex | | | | | |
| Carina | 283.8 | 281.8 | 2.0 | 140 at 4.0 kpc | 8.0 |
| Crux | 309.9 | 309.5 | 0.4 | 042 at 6.0 kpc | 6.3 |
| Norma | 326.4 | 328.4 | 2.0 | 262 at 7.5 kpc | 4.5 |
| Perseus Start | 337.2 | 337.0 | 0.2 | 026 at 7.5 kpc | 3.2 |
| Scutum | 031.3 | 032.4 | 1.1 | 106 at 5.5 kpc | 4.2 |
| Sagittarius | 050.7 | 050.7 | 0.0 | 0 at 3.3 kpc | 6.3 |

Notes:

(a) Y is chosen at or near the 'Potential Minimum' of the density wave, where a broad diffuse CO J=1-0 is peaking in intensity when a radio telescope scans in Galactic Longitude with a broad beamwidth (typically 8.8 arcmin).

(b) Original data for 870μm dust from Table 1 in Vallée 2016a, for masers from Table 1 in Vallée 2020a, for HII complexes from Table 3 in Vallée 2016a. Published updates since were included.

- - - - - - - - - - - - -